\documentclass[preprint,preprintnumbers,amsmath,amssymb]{revtex4}

\usepackage{amssymb,latexsym,amsmath}

\usepackage[activeacute,english]{babel}
\usepackage{fancyhdr}
\usepackage{epsfig}
\usepackage{textcomp}
\usepackage{amsfonts}
\usepackage{graphicx}
\usepackage{graphics}

\newcommand{\al}{\alpha}
\newcommand{\pa}{\partial}
\newcommand{\ep}{\epsilon}

\newcommand{\ta}{\tau}

\newcommand{\om}{\omega}

\newcommand{\de}{\delta}

\begin{document}

\title[long]{$gl_{n+1}$ algebra of Matrix Differential Operators and Matrix
Quasi-exactly-solvable Problems}

\author{Yu.F.~Smirnov (deceased)}
%\footnote{deceased}
%
\author{A.V.~Turbiner}
\email{turbiner@nucleares.unam.mx}
\affiliation{Instituto de Ciencias Nucleares, Universidad Nacional
Aut\'onoma de M\'exico, Apartado Postal 70-543, 04510 M\'exico,
D.F., Mexico}

\date{June 6, 2013}

\begin{abstract}
The generators of the algebra $gl_{n+1}$ in a form of differential operators of the first order acting on ${\bf R}^n$ with matrix coefficients are explicitly written. The algebraic Hamiltonians
for a matrix generalization of $3-$body Calogero and Sutherland models are presented.
\end{abstract}

\maketitle

\begin{center}
 {\it Dedicated to Miloslav Havlicek on the occasion of his 75th birthday}
% \\ (invited contribution)}
\end{center}

\section*{Introduction}

This work has a certain history related to Miloslav Havlicek. Due to an important occasion
of Miloslav's 75th birthday we think this story has to be revealed.
About 25 years ago when quasi-exactly-solvable Schroedinger equations with the hidden
algebra $sl_2$ were discovered \cite{Turbiner:1988} one of the present authors (AVT)
approached to Israel M Gelfand and asked about the existence of the algebra $gl_{n+1}$
of the matrix differential operators. Instead of giving an answer Israel Moiseevich
said that M Havlicek knows the answer and he must be asked. A set of the Dubna preprints
was given (see \cite{Burdik:1985}-\cite{Burdik:2004} and reference therein). Then AVT studied
them for many years firstly separately and then together with YuFS who also happened to have
the same set of preprints. The results of those studies are presented below. Doing these
studies we always kept in mind that the constructive answer exists and is known
to Miloslav. Thus, we are certain, at least, some of presented results are known to Miloslav.
We were simply unable to understand and then indicate where they can be found.
Our main goal is to find a mixed representation of the algebra $gl_{n+1}$ which
contains both matrices and differential operators in a non-trivial
way. Then to generalize it to a polynomial algebra which we call $g^{(m)}$ (see below, Section 4). 
While the another goal is to apply obtained representations for a construction of the algebraic forms of (quasi)-exactly-solvable matrix Hamiltonians.

\section{The algebra $gl_{n+1}$ in mixed representation}

%\vskip 1cm

%\hskip 1cm
Let us take the algebra $gl_{n}$ and consider the vector field
representation
\begin{equation}
\label{egln}
  {\tilde E}_{ij}\ =\ x_i \pa_j\ ,\ i,j=1,\ldots n \quad ,\
  x \in {\bf R}^n \ .
\end{equation}
It obeys the canonical commutation relations
\begin{equation}
\label{comm-e1}
  [{\tilde E}_{ij} , {\tilde E}_{kl}]\ =\ \de_{j k} {\tilde E}_{il}-\
  \de_{il} {\tilde E}_{kj} \ .
\end{equation}
On the other hand, let us consider another representation $M_{pm}\ ,\ p,m =1,\ldots n$
of the algebra $gl_{n}$ in terms of some operators (matrix, finite-difference
etc) with a condition that all `cross-commutators' between these two representations vanish
\begin{equation}
\label{e2}
  [{\tilde E}_{ij} , M_{pm}]\ =\ 0\ .
\end{equation}
Let us choose $M_{pm}$ to obey canonical commutation relations
\begin{equation}
\label{e2-1}
  [M_{ij} , M_{kl}]\ =\ \de_{j k} M_{il}-\ \de_{il} M_{kj} \ ,
\end{equation}
(cf. (\ref{comm-e1})). It is evident that the sum of these two representations is also representation,
\begin{equation}
\label{eij}
E_{ij} \equiv {\tilde E}_{ij} + M_{ij} \in gl_n\ .
\end{equation}

Now we consider an embedding of $gl_n \subset gl_{n+1}$ trying to
complement the representation (\ref{egln}) of the algebra $gl_n$
up to the representation of the algebra $gl_{n+1}$. In principle,
it can be done due to the existence of the Weyl-Cartan decomposition,
$$gl_{n+1} = L \ {\oplus} \ (gl_{n}\oplus {\bf I})\ \oplus\ U$$
with the property
\begin{equation}
\label{WC}
        gl_{n+1} = L \ \rtimes \ (gl_{n}\oplus {\bf I})\ \ltimes\ U \ ,
\end{equation}
where $L (U)$ is the commutative algebra of the lowering (raising) generators with a property
$[L, U]\ =\ gl_{n}\oplus {\bf I}$. Thus, it realizes a property of the Gauss decomposition
of $gl_{n+1}$.
It is worth emphasizing that the $\dim (L) = \dim (U)= n$.

Obviously, the lowering generators (of negative grading) from $L$ can be given by derivations
\begin{equation}
\label{e-}
  T_i^- \ =\ \pa_i\ ,\ i=1,\ldots, n \quad ,\quad \pa_i \equiv \frac{\pa}{\pa x_i}\ ,
\end{equation}
(see e.g. \cite{RT:1995}) when assuming that all commutators
\begin{equation}
\label{e2-comm}
  [T_{i}^- , M_{pm}]\ =\ 0\ ,
\end{equation}
vanish. Likely, it implies that the only possible choice for
$M_{pm}$ exists when they are given either by matrices or act in a space which is a
complement to the $x \in {\bf R}^n$. It is easy to check that
%\[
%[{\tilde E}+M, T^-]\ =\ T^-\quad .
%\]

\[
[E_{ij}, T^-_k]\ =\ -\de_{ik} T^-_j \quad .
\]
Now we have to add the Euler-Cartan generator of the $gl_{n}$ algebra, see (\ref{WC})
\begin{equation}
\label{e0}
  - E_0\ =\ \sum_{j=1}^n x_j \pa_j - k\ ,
\end{equation}
where $k$ is arbitrary constant. Raising generators from $U$ are chosen as
\[
 - T_i^+ \ =\ -x_i E_0 + \sum_{j=1}^n x_j M_{ij}\ 
\]
\begin{equation}
\label{e+}
 =\ x_i (\sum_{j=1}^n x_j \pa_j - k) +
 \sum_{j=1}^n x_j M_{ij} \ ,\ i=1,\ldots, n \ .
\end{equation}
(cf. e.g. \cite{RT:1995}).
Needless to say that one can check explicitly that $T_i^-,\ E_{ij},\ E_0,\ T_i^+$ span
the algebra $gl_{n+1}$. In particular,
\[
[E, T^+]\ =\ T^+ \quad ,
\]
and
\[
[T^+_i, T^-_j]\ =\ E_{ii}-\de_{ij} E_0 \quad .
\]

If the parameter $k$ takes non-negative integer the algebra $gl_{n+1}$ spanned
by the generators (\ref{eij}), (\ref{e-}), (\ref{e0}) - (\ref{e+}) appears in
finite-dimensional representation. There exists a linear finite-dimensional space
of polynomials of a finite-order in the space of columns/spinors of a finite length
which is common invariant subspace for all generators (\ref{eij}), (\ref{e-}), (\ref{e0}) - (\ref{e+}). This finite-dimensional representation is irreducible.

The non-negative integer parameter $k$ has a meaning of the length
of the first row of the Young tableau of $gl_{n+1}$, describing the totally
symmetric representation (see below). All other parameters are coded in
$M_{ij}$, which corresponds to arbitrary Young tableau of $gl_n$.
Thus, we have some peculiar splitting of the Young tableau.

Each representation is characterized by Gelfand-Tseitlin signature,
$[m_{1,n}, \ldots m_{nn}]$, where $m_{in} \geq m_{i+1,n}$ and
their difference is positive integer. Each basic vector is
characterized by the Gelfand-Tseitlin scheme.
Explicit form of the representation is given by Gelfand-Tseitlin formulas
\cite{GT:1950}.

It can be demonstrated that all Casimir operators of $gl_{n+1}$ in this
realization (\ref{eij}), (\ref{e-}), (\ref{e0}) - (\ref{e+}) are expressed in $M_{ij}$,
thus, do not depend on $x$. They coincide to the Casimir operators of the $gl_n$-subalgebra
realized by matrices $M_{ij}$.

\section{Example: the algebra $gl_3$ in mixed representation}

In the case of the algebra $gl_3$ the generators (\ref{eij}), (\ref{e-}), (\ref{e0}) - (\ref{e+}) take the form
\[
E_{11}=x_1\pa_1 + M_{11}\ ,\ E_{22}=x_2\pa_2 + M_{22}\ ,
\]
\[
E_{12}=x_1\pa_2 + M_{12}\ ,\ E_{21}=x_2\pa_1 + M_{21}\ ,
\]
\[
E_{0}=k - x_1\pa_1 - x_2\pa_2 \ ,
\]
\[
T_{1}^-=\pa_1 \ ,\ T_{2}^-=\pa_2 \ ,
\]
\[
T_{1}^+=x_1(k - x_1\pa_1 - x_2\pa_2) - x_1 M_{11}- x_2 M_{12}\ ,
\]
\begin{equation}
\label{gl3}
  T_{2}^+=x_2(k - x_1\pa_1 - x_2\pa_2) - x_1 M_{21}- x_2 M_{22}\ .
\end{equation}
The Casimir operators of $gl_3$ in this realization are given by
\[
C_1 = E_{11}+E_{22}+E_0 = k + M_{11}+M_{22}= k + C_1(M)\ ,
\]
\[
C_2 = E_{12}E_{21}+E_{21}E_{12}+T_{1}^+T_{1}^-+T_{1}^-T_{1}^+
+T_{2}^+T_{2}^- + T_{2}^-T_{2}^+ 
\]
\[
+ E_{11}^2 + E_{22}^2+E_0^2
\]
\[
  = k(k+2) + M_{11}^2+M_{22}^2 +
 M_{12}M_{21} + M_{21}M_{12} - M_{11} - M_{22}
\]
\[ 
 =k(k+2)+C_2(M)-C_1(M)\ ,
\]
and, finally,
\[
C_3= -\frac{1}{2}C_1^3+\frac{3}{2}C_1C_2+3C_2-2 C_1^2-2C_1\ .
\]
In this realization the Casimir operator $C_3$ is algebraically dependent
on $C_1$ and $C_2$. In fact, $C_1$ and $C_2$ are nothing but the Casimir operators of
the $gl_2$ sub-algebra.
Therefore, the center of the $gl_3$ universal enveloping algebra
in the realization (\ref{gl3}) is generated by the Casimir operators of the $gl_2$
sub-algebra realized by $M_{ij}$. Thus, it seems natural that these reps are irreducible.

Now we consider concrete matrix realizations of the $gl_2$-subalgebra in our scheme.

{\bf (0)}. Reps in $1 \times 1$ matrices. It corresponds to the trivial
representation of $gl_2$,
\[
 M_{11}=M_{12}=M_{21}=M_{22}=0\ .
\]
This is $[k,0]$ or, in other words, the symmetric representation (the Young tableau
has two rows of the length $k$ and 0, correspondingly). We also can call it a {\it scalar} representation, since the generators
\[
E_{11}=x_1\pa_1 \ ,\ E_{22}=x_2\pa_2 \ ,
\]
\[
E_{12}=x_1\pa_2 \ ,\ E_{21}=x_2\pa_1 \ ,
\]
\[
E_{0}=k - x_1\pa_1 - x_2\pa_2 \ ,
\]
\[
T_{1}^-=\pa_1 \ ,\ T_{2}^-=\pa_2 \ ,
\]
\[
T_{1}^+=x_1(k - x_1\pa_1 - x_2\pa_2) \ ,
\]
\begin{equation}
\label{gl3-0}
T_{2}^+=x_2(k - x_1\pa_1 - x_2\pa_2) \ ,
\end{equation}
act on one-component spinors or, in other words, on scalar functions (see e.g. \cite{RT:1995}). The Casimir operators are:
\[
C_1= k\ ,
\]
\[
C_2= k(k+2)\ .
\]
If the parameter $k$ takes non-negative integer the algebra $gl_{3}$ spanned
by the generators (\ref{gl3-0}) appears in finite-dimensional representation. Its finite-dimensional representation space is a space of polynomials
\begin{equation}
\label{P2-0}
 {\cal P}_{k,0}\ =\ \langle {x_{1}}^{p_1}
 {x_{2}}^{p_2} \vert \ 0 \le p_1 + p_2 \le k \rangle\ ,\ k=0,1,2,\ldots\
\end{equation}

Namely in this representation (\ref{gl3-0}) the algebra $gl_3$ appears as the hidden algebra of the 3-body Calogero and Sutherland models \cite{RT:1995}, $BC_2$ rational and trigonometric, and $G_2$ rational models \cite{Turbiner:1998} and even of the $BC_2$ elliptic model \cite{Turbiner:2012tt}.

{\bf (I)}. Reps in $2 \times 2$ matrices.

Take $gl_2$ in two-dimensional reps by $2 \times 2$ matrices,
\[
M_{11}\ =\ \left( \begin{array}{cc} 1 & 0
\\ 0 & 0
\end{array}  \right)\ ,\ M_{22}\ =\ \left( \begin{array}{cc} 0 & 0
\\ 0 & 1
\end{array}  \right)\ ,
\]
\[
M_{12}\ =\ \left( \begin{array}{cc} 0 & 1
\\ 0 & 0
\end{array}  \right)\ ,\ M_{21}\ =\ \left( \begin{array}{cc} 0 & 0
\\ 1 & 0
\end{array}  \right)\ ,
\]
Then the generators (\ref{gl3}) of $gl_3$ are:
\begin{equation*}
\label{T-,2}
 T^-_{1}\ =\ \left( \begin{array}{cc} \pa_1 & 0
\\ 0 & \pa_1
\end{array}  \right)\ ,
\ T^-_{2} \ =\ \left( \begin{array}{cc} \pa_2 & 0
\\ 0 & \pa_2
\end{array}  \right)\ ,
\end{equation*}
\begin{equation*}
\label{E1,2}
 E_{11}=\left( \begin{array}{cc} x_1\pa_1 +1 & 0
\\ 0 & x_1\pa_1
\end{array}  \right),
\ E_{12}=\left( \begin{array}{cc} x_1 \pa_2 & 1
\\ 0 & x_1 \pa_2
\end{array}  \right),
\end{equation*}
\begin{equation*}
\label{E2,2}
 E_{21}=\left( \begin{array}{cc} x_2 \pa_1 & 0
\\ 1 & x_2 \pa_1
\end{array}  \right),
 E_{22}=\left( \begin{array}{cc} x_2\pa_2  & 0
\\ 0 & x_2\pa_2+1
\end{array}  \right),
\end{equation*}
\begin{equation*}
\label{E,2}
 E_{0}\ =\ \left( \begin{array}{cc} k -x_1\pa_1 - x_2\pa_2  & 0
\\ 0 & k - x_1\pa_1 - x_2\pa_2
\end{array}  \right)\ ,
\end{equation*}
\begin{equation*}
\label{T1+,2}
 T^+_{1}=\left( \begin{array}{cc} x_1(k-1 -x_1\pa_1 - x_2\pa_2)  &
 -x_2
\\ 0 & x_1(k - x_1\pa_1 - x_2\pa_2)
\end{array}  \right),
\end{equation*}
\begin{equation}
\label{gl3-1}
 T^+_{2}=\left( \begin{array}{cc} x_2(k -x_1\pa_1 - x_2\pa_2)  & 0
\\ -x_1 & x_2(k-1 - x_1\pa_1 - x_2\pa_2)
\end{array}  \right).
\end{equation}

This is $[k,1]-$representation (the Young tableau has two rows of
the length $k$ and 1, correspondingly) and their Casimir operators
are:
\[
C_1= k+1\ ,
\]
\[
C_2= (k+1)^2\ .
\]
If the parameter $k$ takes non-negative integer the algebra $gl_{3}$ spanned
by the generators (\ref{gl3-1}) appears in finite-dimensional representation. 

Let us consider several different values of $k$ in detail.

{\bf (i)}\ $k=1$. Then three-dimensional representation space
$V_1^{(2)}$ appears to be spanned by:
\begin{equation}
\label{k=1}
 P_-={0 \brack 1}\ ,\ P_+={1 \brack 0}\ ,\
 Y_1={x_2 \brack -x_1}\ .
\end{equation}
It corresponds to antiquark multiplet in standard (fundamental)
representation. The Newton polygon is triangle with points
$P_{\pm}$ as vortices at base.

{\bf (ii)} $k=2$. Then eight-dimensional representation  space
$V_2^{(2)}$ appears to be spanned by:
\[
 P_- = {0 \brack 1}\ ,\ P_+ = {1 \brack 0}\ ,
\]
\[
 P^{(1)}_- = {0 \brack x_2},\ Y^{(1)}_{1} = {0 \brack x_1}, \ 
 Y^{(2)}_{1} = {x_2 \brack 0},\ P^{(1)}_+ = {x_1 \brack 0},
\]
\begin{equation}
\label{k=2}
 Y_2 = {x_2^2 \brack -x_1 x_2}\ ,
 \ Y_3 = {x_1 x_2 \brack -x_1^2}\ .
\end{equation}
It corresponds to octet in standard (fundamental) representation.
The space $V_2^{(2)}$ contains $V_1^{(2)}$ as a subspace,
$V_1^{(2)} \subset V_2^{(2)}$. It should be mentioned that
$Y_1=-Y^{(1)}_{1}+Y^{(2)}_{1}$. Now the Newton polygon is the
hexagon where the central point is doubled being presented by
$Y^{(1,2)}_{1}$ and lower (upper) base has a length two being
given by $P_{\pm}$ ($Y_{2,3}$).

\renewcommand{\theequation}{17.{\arabic{equation}}}
\setcounter{equation}{0}

{\bf (iii)} $k=3$. The representation space $V_3^{(2)}$ is
15-dimensional. In addition to $P_{\pm}, P^{(1)}_{\pm}$ and
$Y^{(1,2)}_1$ (see (\ref{k=1}) and (\ref{k=2})) it contains several vectors
more, namely,
\begin{equation}
\label{k=3.1}
 P^{(2)}_- = {0 \brack  x_2^2}\ ,
 \ P^{(2)}_+ = {x_1^2 \brack 0}\ ,
\end{equation}
which are situated on the $\pm$-sides of the Newton hexagon,
doubling of the points corresponding to $Y_{2,3}$ (see (\ref{k=2}))
\[
 Y^{(1)}_2 = {0 \brack x_1 x_2}\ ,\
 Y^{(2)}_2 = {x_2^2 \brack 0}\ ,
\]
\begin{equation}
\label{k=3.2}
 Y^{(1)}_3 = {0 \brack x_1^2}\ ,\
 Y^{(2)}_3 = {x_1 x_2 \brack 0}\ ,
\end{equation}
and plus extra three vectors on the boundary
\begin{equation}
\label{k=3.3}
 Y_8 = {x_2^3 \brack -x_1x_2^2}\ ,\ Y_9 = {x_1 x_2^2 \brack -x_1^2 x_2}\ ,
 \ Y_{10} = {x_1^2 x_2 \brack -x_1^3}\ .
\end{equation}
It is clear that $V_1^{(2)} \subset V_2^{(2)} \subset V_3^{(2)}$.
All internal points of the Newton hexagon are double points, while
the points on the boundary are single ones.

\renewcommand{\theequation}{{\arabic{equation}}}
\setcounter{equation}{17}

{\bf (iv)} $\forall k$. The finite-dimensional representation space $V_k^{(2)}$
has dimension $k(k+2)$ and is presented by the Newton hexagon which contains
$(k+1)$ horizontal layers, lower base has length two, while the
upper one has length $k$ (see Fig.1 as an illustration for $k=4$). All
internal points of the Newton hexagon are double points, while the
points on the boundary are single ones.
\begin{figure}
\begin{center}
%    \fbox{
     %\psfig{figure=K_2.ps,width=5.in,angle=-90}
     \includegraphics*[width=3.in,angle=-90]{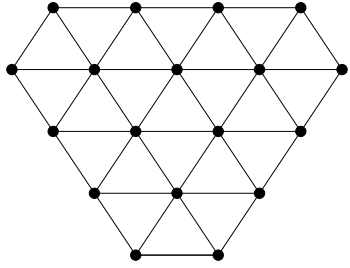}
     \caption{Newton hexagon for the representation space $V_4^{(2)}$
     of the $[4,1]$ representation of dimension 24. }
    \label{fig:1}
\end{center}
\end{figure}
 Except for $k$ vectors of the last (highest) layer of
the Newton hexagon, the remaining $k(k+1)$ vectors span the space
of all possible two-component spinors with components given by
inhomogeneous polynomials in $x_1,x_2$ of degree not higher than
$(k-1)$. We denote this space as ${\tilde V}_k^{(2)} \subset
V_k^{(2)}$. The non-trivial task is to describe $k$ vectors of the
last (highest) layer of the hexagon. After some analysis one can
find out that they have a form
\begin{equation}
\label{k}
Y_{k(k+1)+i} = {x_2^{k-i}x_1^i \brack - x_2^{k-i-1} x_1^{i+1}}\ ,\
i=0,1,2,\ldots (k-1)\ ,
\end{equation}
hence span a non-trivial $k$-dimensional subspace of spinors with
components given by specific homogeneous polynomials of degree
$k$.

{\bf (II)}\ Reps in $3 \times 3$ matrices.

Take $gl_2$ in three-dimensional reps by $3 \times 3$ matrices,
\[
M_{11}=\left( \begin{array}{ccc} 2 & 0 & 0
\\ 0 & 1 & 0
\\ 0 & 0 & 0
\end{array}  \right),
M_{22}=\left( \begin{array}{ccc} 0 & 0 & 0
\\ 0 & 1 & 0
\\ 0 & 0 & 2
\end{array}  \right),
\]
\[
M_{12}=\left( \begin{array}{ccc} 0 & \sqrt{2} & 0
\\ 0 & 0 & \sqrt{2}
\\ 0 & 0 & 0
\end{array}  \right),
M_{21}=\left( \begin{array}{ccc} 0 & 0 & 0
\\ \sqrt{2} & 0 & 0
\\ 0 & \sqrt{2}  & 0
\end{array}  \right).
\]
Then the generators (\ref{gl3}) of $gl_3$ are:
\begin{equation}
\label{T-,3}
 T^-_{1}=\left( \begin{array}{ccc} \pa_1 & 0 & 0
\\ 0 & \pa_1 & 0
\\ 0 & 0 & \pa_1
\end{array}  \right),
\ T^-_{2}=\left( \begin{array}{ccc} \pa_2 & 0 & 0
\\ 0 & \pa_2 & 0
\\ 0 & 0 & \pa_2
\end{array}  \right),
\end{equation}
\[
\label{E1,3}
 E_{11}\ =\ \left( \begin{array}{ccc} x_1\pa_1+2 & 0 & 0
\\ 0 & x_1\pa_1+1 & 0
\\ 0 & 0 & x_1\pa_1
\end{array}  \right)\ ,
\]
\[
\ E_{12} \ =\ \left( \begin{array}{ccc} x_1\pa_2 & \sqrt{2} & 0
\\ 0 & x_1\pa_2 & \sqrt{2}
\\ 0 & 0 & x_1\pa_2
\end{array}  \right)\ ,
\]
\[
\ E_{21} \ =\ \left( \begin{array}{ccc} x_2\pa_1 & 0 & 0
\\ \sqrt{2} & x_2\pa_1 & 0
\\ 0 & \sqrt{2} & x_2\pa_1
\end{array}  \right)\ ,
\]
\[
\label{E2,3}
 E_{22}\ =\ \left( \begin{array}{ccc} x_2\pa_2 & 0 & 0
\\ 0 & x_2\pa_2+1 & 0
\\ 0 & 0 & x_2\pa_2+2
\end{array}  \right)\ ,
\]
\[
\label{E,3}
 E_{0}=\left( \begin{array}{ccc} k -x_1\pa_1 - x_2\pa_2 & 0 & 0
\\ 0 & k - x_1\pa_1 - x_2\pa_2 & 0
\\ 0 & 0 & k - x_1\pa_1 - x_2\pa_2
\end{array}  \right),
\]
\[
\label{T1+,3}
 T^+_{1}=\left(
 \begin{array}{ccc} x_1(k-2 -x_1\pa_1 - x_2\pa_2)  & -\sqrt{2}x_2 & 0
\\ 0 & x_1(k - 1 - x_1\pa_1 - x_2\pa_2) & -\sqrt{2}x_2
\\ 0 & 0 & x_1(k - x_1\pa_1 - x_2\pa_2)
\end{array}  \right),
\]
\[
\label{T2+,3}
 T^+_{2}=\left(
 \begin{array}{ccc} x_2(k -x_1\pa_1 - x_2\pa_2)  & 0 & 0
\\ -\sqrt{2}x_1 & x_2(k - 1 - x_1\pa_1 - x_2\pa_2) & 0
\\ 0 & -\sqrt{2}x_1 & x_2(k - 2 - x_1\pa_1 - x_2\pa_2)
\end{array}  \right).
\]
This is $[k,2]-$representation (Young tableau has two rows of the
length $k$ and 2, correspondingly) and their Casimir operators
are:
\[
C_1= k+2\ ,
\]
\[
C_2= (k+1)^2+3\ .
\]

As an illustration let us show explicitly finite-dimensional representation spaces for $k=2,3$.

{\bf (i)}\ $k=2$. Then six-dimensional representation space $V_2^{(3)}$ appears to be spanned by:
\begin{equation}
\label{k=2;3}
 P_-=\left( \begin{array}{c} 0 \\ 0 \\ 1 \end{array}  \right)\ ,
 \ P_0=\left( \begin{array}{c} 0 \\ 1 \\ 0 \end{array}  \right)\ ,
 \ P_+=\left( \begin{array}{c} 1 \\ 0 \\ 0 \end{array}  \right)\ ,
\end{equation}
\[
 Y_1=\left( \begin{array}{c} 0 \\ x_2 \\ -\sqrt{2}x_1 \end{array} \right)\ ,
 \ Y_2=\left( \begin{array}{c} -\sqrt{2}x_2 \\ x_1 \\ 0 \end{array} \right)\ ,
 \ Y_3=\left( \begin{array}{c} x_2^2 \\ -\sqrt{2}x_1 x_2 \\ x_1^2
 \end{array}  \right)\ .
\]
It corresponds to `di-antiquark' multiplet.

{\bf (ii)} $k=3$. 

Then 15-dimensional representation space $V_3^{(3)}$ appears to be spanned by:

\begin{equation}
\label{k=3;3}
 P_-=\left( \begin{array}{c} 0 \\ 0 \\ 1 \end{array}  \right)\ ,
 \ P_0=\left( \begin{array}{c} 0 \\ 1 \\ 0 \end{array}  \right)\ ,
 \ P_+=\left( \begin{array}{c} 1 \\ 0 \\ 0 \end{array}  \right)\ ,
\end{equation}
\[
 Y^{(1)}_1=\left( \begin{array}{c} 0 \\ x_2 \\ 0 \end{array} \right)\ ,
 \ Y^{(2)}_1=\left( \begin{array}{c} 0 \\ 0 \\ x_1 \end{array} \right)\ ,
 \ Y^{(1)}_2=\left( \begin{array}{c} x_2 \\ 0 \\ 0 \end{array} \right)\ ,
 \ Y^{(2)}_2=\left( \begin{array}{c} 0 \\ x_1 \\ 0 \end{array} \right)\ ,
\]
\[
 P^{(1)}_-=\left( \begin{array}{c} 0 \\ 0 \\ x_2 \end{array} \right)\ ,
 \ P^{(1)}_+=\left( \begin{array}{c} x_1 \\ 0 \\ 0 \end{array} \right)\ ,
\]
\[
 \ Y^{(1)}_3=\left( \begin{array}{c}  -\sqrt{2}x_2^2 \\ x_1 x_2 \\ 0
  \end{array}  \right)\ ,
 \ Y^{(2)}_3=\left( \begin{array}{c} 0\\ x_1 x_2 \\ -\sqrt{2}x_1^2
  \end{array}  \right)\ ,
 \ Y_4=\left( \begin{array}{c} 0 \\ -\sqrt{2}x_2^2 \\ 2x_1 x_2
  \end{array}  \right)\ ,
 \ Y_5=\left( \begin{array}{c} 2x_1 x_2 \\ -\sqrt{2}x_1^2 \\ 0
  \end{array}  \right)\ ,
\]

\[
 Y_6=\left( \begin{array}{c} x_2^3 \\ -\sqrt{2}x_1 x_2^2 \\ x_1^2 x_2
 \end{array} \right)\ ,
 \ Y_7=\left( \begin{array}{c} x_1x_2^2 \\ -\sqrt{2}x_1^2 x_2 \\ x_1^3
 \end{array} \right)\ .
\]

\begin{figure*}
\begin{center}
%    \fbox{
     %\psfig{figure=smirnovA.ps,width=5.in,angle=-90}
     \includegraphics*[width=0.6\linewidth,angle=-90]{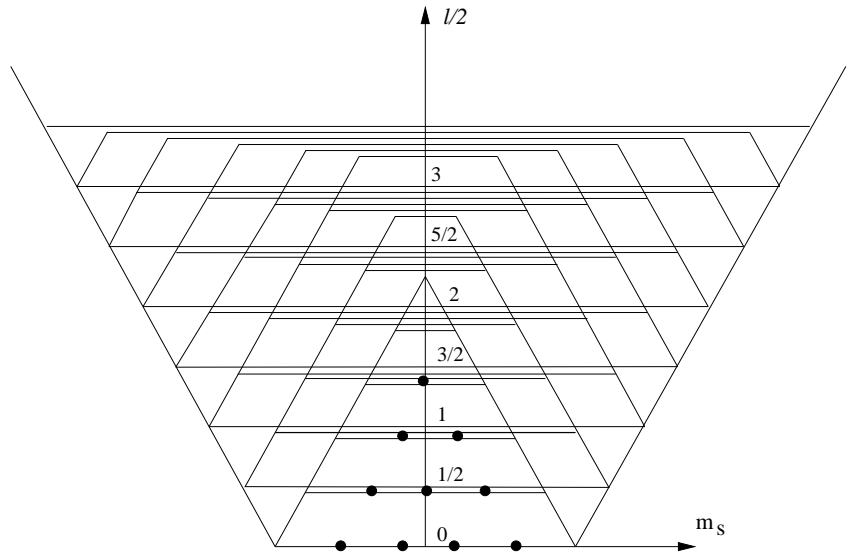}
     \caption{Verma module with lowest weight (057)) for $s=5/2$}
    \label{fig:2}
\end{center}
\end{figure*}
\clearpage

It is worth mentioning that as a consequence of a particular realization of the generators (\ref{gl3}) of the $gl_3$ algebra there exist a certain relations between generators other
than ones given by the Casimir operators.
%The simplest example is
%\[
% E_{11} + E_{22} + E_0 = k+n
%\]
The first observation is that there are no linear relations between generators of such
a type. Some time ago there were found nine quadratic relations between $gl_3$ generators
taken in scalar representation (\ref{gl3-0}) other than Casimir operators \cite{Turbiner:1994}. Surprisingly, a certain modifications of these relations also exist for $[kn]$
mixed representations (\ref{gl3}),
\renewcommand{\theequation}{Art.{\arabic{equation}}}
\setcounter{equation}{0}
\begin{equation}
\label{a=1}
 - T^+_1 E_{22}\ +\ T^+_2 E_{12} \ = \
\end{equation}
\[
x_1[M_{22}x_1 \pa_1+ M_{11} x_2 \pa_2+(M_{11}-k) M_{22} - M_{21}
M_{12}]
\]
\[
-x_2 (x_1 \pa_1-k-1)M_{12}\ -\ M_{21} x_1^2\pa_2 \ \equiv \
-\tilde T^+_1\ ,
\]
\begin{equation}
\label{a=2}
 - T^+_2 E_{11}\ +\ T^+_1 E_{21} \ = \
\end{equation}
\[
x_2[M_{22}x_1 \pa_1+ M_{11} x_2 \pa_2+(M_{22}-k) M_{11} - M_{12}
M_{21}]
\]
\[
-x_1 (x_2 \pa_2-k-1)M_{21}\ -\ M_{12} x_2^2\pa_1 \ \equiv \
-\tilde T^+_2\ ,
\]
\begin{equation}
\label{a=3}
 -E_{12} (E_0+1)\ +\ T^+_1 T^-_2\ =
\end{equation}
\[
 M_{12} (x_1 \pa_1 -k-1) -  M_{11} x_1 \pa_2 \ \equiv \
-\tilde E_{12}\ ,
\]
\begin{equation}
\label{a=4}
 -E_{21} (E_0+1)\ +\ T^+_2 T^-_1\ =
\end{equation}
\[
 M_{21} (x_2 \pa_2 -k-1) - M_{22} x_2 \pa_1 \ \equiv \
-\tilde E_{21}\ ,
\]
\begin{equation}
\label{a=5}
 T^+_1 T^-_1\  - E_{11} (1+E_0)\ =
\end{equation}
\[
 M_{11} x_2 \pa_2
 - M_{12} x_2 \pa_1 - (k+1) M_{11} \ \equiv \
-\tilde E_{11}\ ,
\]
\begin{equation}
\label{a=6}
 T^+_2 T^-_2\  - E_{22} (1+E_0)\ =
\end{equation}
\[
 M_{22} x_1 \pa_1
 - M_{21} x_1 \pa_2 - (k+1) M_{22} \ \equiv \
-\tilde E_{22}\ ,
\]
\begin{equation}
\label{a=7}
 E_{12}E_{21} - E_{11}E_{22}- E_{11}\ =
\end{equation}
\[
 M_{12} x_2 \pa_1 + M_{21} x_1 \pa_2
 -M_{22} x_1 \pa_1 - M_{11} x_2 \pa_2 +
 M_{12}M_{21}
\]
\[ 
 -M_{11}M_{22}-M_{11}\ \equiv \
-\hat E_{11}\ ,
\]
\begin{equation}
\label{a=8}
 E_{22}T^-_{1} - E_{21}T^-_{2} = M_{22}\pa_1 - M_{21}\pa_2\ \equiv \
 \tilde T^-_{1}\ ,
\end{equation}
\begin{equation}
\label{a=9}
 E_{12}T^-_{1} - E_{11}T^-_{2} = M_{12}\pa_1 - M_{11}\pa_2\ \equiv \
 -\tilde T^-_{2}\ .
\end{equation}
Not all these relations are independent. It can be shown that one
relation is linearly dependent 
since the sum (Art.5)+(Art.6)+(Art.7) gives the second Casimir operator 
$C_2$.

At least, in scalar case we can assign a natural (vectorial)
grading to generators. Above relations reflect a certain
decomposition of gradings as well,
\[
(1,0) (0,0)\ =\ (0,1) (1,-1)
\]
\[
(0,1) (0,0)\ =\ (1,0) (-1,0)
\]
for the first two relations,
\[
(1,-1) (0,0)\ =\ (1,0) (0,-1)
\]
\[
(-1,1) (0,0)\ =\ (0,1) (-1,0)
\]
for the second two,
\[
(1,0) (-1,0)\ =\ (0,0) (0,0)
\]
\[
(0,1) (0,-1)\ =\ (0,0) (0,0)
\]
\[
(1,-1) (-1,1)\ =\ (0,0) (0,0)
\]
for three before the last two
\[
(0,0) (-1,0)\ =\ (-1,1) (0,-1)
\]
\[
(0,0) (0,-1)\ =\ (1,-1) (-1,0)
\]
for last two.

\renewcommand{\theequation}{{\arabic{equation}}}
\setcounter{equation}{21}

\section{The algebra $g^{(m)}$ in mixed representation}

The basic property which was used to construct mixed representation of the algebra $gl_{n+1}$
is the existence of the Weyl-Cartan decomposition $gl_{n+1} = L \ {\oplus} \ (gl_{n}\oplus {\bf I})\ \oplus\ U$ with the property (\ref{WC}). One can pose a question about
the existence of other algebras than $gl_{n+1}$ for which the Weyl-Cartan decomposition with the property (\ref{WC}) holds. The answer is affirmative. Let us consider the important particular case of the Cartan algebra $gl_{2}\oplus {\bf I}$ and construct a realization of a new algebra denoted $g^{(m)}$ with the property
\begin{equation}
\label{WC-2}
    g^{(m)}\ =\ L_{m+1} \rtimes \ (gl_{2}\oplus {\bf I})\ \ltimes\ U_{m+1}
    \ ,
\end{equation}
where $L_m(U_m)$ is the commutative algebra of the lowering (raising) generators with a property
$[L_m, U_m]\ =\ P_{m-1}(gl_{2}\oplus {\bf I})$ with $P_{m-1}$ as a polynomial of degree $(m-1)$ in generators of $gl_{2}\oplus {\bf I}$. Thus, it realizes a property of the generalized Gauss decomposition. The emerging algebra is a polynomial algebra.
It is worth emphasizing that the realization we are going to construct appears at $\dim (L_k) = \dim (U_m)= m$. For $m=1$ the algebra $g^{(1)} = gl_3$, see (\ref{WC}). Our final goal to build the realization of (\ref{WC-2}) in terms of finite order differential operators acting on the plane ${\bf R}^2$.

The simplest realization of the algebra $gl_2$ by differential operators in two variables is the vector field representation, see (\ref{egln}) at $n=2$. It was exactly this representation which was used to construct the representation of $gl_3$ acting of ${\bf R}^2$, see (\ref{gl3}), (\ref{gl3-0}). In this case $\dim (L_m) = \dim (U_m)= 2$. We are unable to find other algebras with $\dim (L_m) = \dim (U_m)> 2$. However, there exists another representation of the algebra $gl_2$ by the first order differential operators in two variables,
\[
{\tilde J}_{12}\  =\  \pa_x\ ,
\]
\[
{\tilde J}^{(k)}_{11}\  =\ -x \pa_x\ +\ \frac{k}{3} \ ,
 \ {\tilde J}^{(k)}_{22}\  =\ -x \pa_x\ +\ s y\pa_y\ ,
\]
\begin{equation}
\label{gl2r}
       {\tilde J}^{(k)}_{21}\  =\ x^2 \pa_x \  +\ s x y \pa_y \ - \ k x\ ,
\end{equation}
(see S.~Lie, \cite{Lie} at $k=0$, W.~Miller \cite{Miller:1968} and A.~Gonz\'alez-Lop\'ez et al, \cite{glko} at $k \neq 0$ (Case 24)), where $s, k$ are arbitrary numbers. These generators obey
the standard commutation relations (\ref{comm-e1}) the algebra $gl_2$ in the vector field representation (\ref{egln}).
It is evident that the sum of the two representations, ${\tilde J}_{ij}$ and the matrix one $M_{ij}$ is also representation,
\begin{equation}
\label{jij}
J_{ij} \equiv {\tilde J}_{ij} + M_{ij} \in gl_2\ .
\end{equation}
(cf. (\ref{eij})). It is worth mentioning that the $gl_2$ algebra commutation relations for $M_{pm}$ are taken in a canonical form (\ref{e2-1}). The unity generator ${\bf I}$ in (\ref{WC-2}) is written in a form of generalized Euler-Cartan operator
\begin{equation}
\label{gl20}
  J^{(k)}_{0}\  =\ x \pa_x \  +\ s y \pa_y \ - \ k \ .
\end{equation}

Now let us assume that $s$ is non-negative integer, $s=m,\ m=0,1,2,\ldots$.
Evidently, the lowering generators (of negative grading) from $L_{m+1}$ can be given by
\begin{equation}
\label{g-}
  T_i^- \ =\ x^i \pa_y\ ,\quad i=0, 1,\ldots, m \quad ,
\end{equation}
forming commutative algebra
\begin{equation}
\label{g2-comm}
  [T_{i}^- , T_{j}^-]\ =\ 0\ .
\end{equation}
(cf. \cite{Lie}, \cite{Miller:1968}, \cite{glko}).
Eventually, the generators of the algebra $(gl_2\oplus {\bf I}) \ltimes L_{m+1}$ take the form
\[
J_{12}\  =\  \pa_x + M_{12}\ ,
\]
\[
J^{(k)}_{11}\  =\ -x \pa_x\ +\ \frac{k}{3} + M_{11} \ ,
 \ J^{(k)}_{22}\  =\ -x \pa_x\ +\ m y\pa_y + M_{22},
\]
\begin{equation}
\label{gl2rM}
       J^{(k)}_{21}\  =\ x^2 \pa_x \  +\ m x y \pa_y \ - \ k x + M_{21}\ ,
\end{equation}
with $J^{(k)}_{0}, T_i^-$ given by (\ref{gl20}), (\ref{g-}), respectively.

Let us consider two particular cases of the general construction of the raising generators for the commutative algebra $U$.

(i)
For the first case we take the trivial matrix representation of the $gl_2$,
\[
 M_{11}=M_{12}=M_{21}=M_{22}=0\ .
\]
One can check that one of the raising generators is given by
\begin{equation}
\label{to}
     U_0 = y\pa_{x}^m\ ,
\end{equation}
while all other raising generators are multiple commutators of $J^{(k)}_{21}$ with $U_0$,
\[
    U_i\ \equiv\ \underbrace{[J^{(k)}_{21},[J^{(k)}_{21},[ \ldots J^{(k)}_{21},T_0]\ldots]}_i \ =
\]
\begin{equation}
\label{T}
    y\pa_{x}^{m-i} J^{(k)}_{0} (J^{(k)}_{0}+1)\ldots (J^{(k)}_{0}+i-1) \ ,
\end{equation}
at $i=1,\ldots m$. All of them are differential operators of the fixed degree $m$. The procedure of construction of the operators $U_i$ has a property of nilpotency:
\[
  U_i = 0\ ,\ i > m \ .
\]
In particular, for $m=1$,
\[
   U_0\ =\ y\pa_{x}\ ,\  U_1\ =\ y J^{(k)}_{0}\ =\ y (x \pa_x + y \pa_y - k) \ .
\]
Inspecting the generators $T^-_{0,1},\ J_{ij},\ J^{(n)},\ U_{0,1}$ one can see that they span the algebra $gl_3$, see (\ref{gl3-0}). Hence, the algebra $g^{(1)} \equiv gl_3$.

If the parameter $k$ takes non-negative integer the algebra $g^{(m)}$ spanned
by the generators (\ref{gl2rM}), (\ref{to}), (\ref{T}) appears in finite-dimensional representation. Its finite-dimensional representation space is a triangular space of polynomials
\begin{equation}
\label{P2-m}
 {\cal P}_{k,0}\ =\ \langle {x}^{p_1}
 {y}^{p_2} \vert \ 0 \le p_1 + m p_2 \le k \rangle\ ,\ k=0,1,2,\ldots\
\end{equation}

Namely in this representation the algebra $g^{(m)}$ appears as the hidden algebra of the 3-body $G_2$ trigonometric model \cite{Turbiner:1998} at $m=2$ and of the so-called TTW model at integer $m$, in particular, of the dihedral $I_2(m)$ rational model \cite{TTW:2009}.

(ii) The second case is a certain evident extension when the generators $M_{ij}$ are of arbitrary matrix representation of the algebra $gl_2$. Raising generators (\ref{to}), (\ref{T}) remain the raising generators even if the Cartan generators are given by (\ref{gl2rM}) with arbitrary $M_{ij} \in gl_2$. However, the algebra is not closed: $[T, U] \neq P(gl_2\oplus {\bf I})$ with $P$ as a polynomial of a finite order. It can be fixed, at least, for the case $m=1$. If $M_{ij}$ are generators of $gl_2$ subalgebra of $gl_3$. By adding to $T, U$ generators (\ref{g2-comm}),  (\ref{to}), (\ref{T}) the appropriate matrix generators from $gl_3$, the algebra gets closed. We end up with the $gl_3$ algebra of matrix differential operators
other than (\ref{gl3}). We are not aware about the solution of this problem for the case of $m \neq 1$ except for the case of trivial matrix representation, see case (i).

\section{Extension of the 3-body Calogero Model}

The first {\it algebraic} form for the 3-body Calogero Hamiltonian \cite{Calogero:1969} appears
after gauge rotation with the ground state function, separation of the center-of-mass and
change the variables to elementary symmetric polynomials of the translationally-symmetric coordinates \cite{RT:1995},
\[
  h_{\rm Cal}=-2\ta_2\pa^2_{\ta_2\ta_2}
  -6\ta_3\pa^2_{\ta_2\ta_3} + \frac{2}{3}\ta_2^2\pa^2_{\ta_3\ta_3}
\]
\begin{equation}
\label{e2.1}  
  -[4\om\ta_2+2(1+3\nu)]\pa_{\ta_2} - 6\om\ta_3\pa_{\ta_3} \ .
\end{equation}
These new coordinates are polynomial invariants of the $A_2$ Weyl group.
Its eigenvalues are
\begin{equation}
\label{e2.1a}
    -\ep_p\ =\ 2\om (2p_1 + 3 p_2)\quad,\quad p_{1,2}=0,1,\ldots
\end{equation}

As is shown in Ruhl and Turbiner \cite{RT:1995}, the operator
(\ref{e2.1}) can be rewritten in a Lie-algebraic form in terms of
the $gl(3)$-algebra generators of the representation $[k,0]$. The
corresponding expression is
\[
 h_{\rm Cal} = -2 E_{11} T_{1}^- -
 6 E_{22} T_1^- + \frac{2}{3} E_{12} E_{12} 
\]
\begin{equation}
\label{e2.2} 
 -4 \om E_{11} - 2(1+3\nu) T_1^- - 6\om E_{22} \ .
\end{equation}
Now we can substitute the generators of the representation $[k,n]$
in the form (\ref{gl3})
\[
  {\tilde h}_{\rm Cal}=-2\ta_2\pa^2_{\ta_2\ta_2}
  -6\ta_3\pa^2_{\ta_2\ta_3} + \frac{2}{3}\ta_2^2\pa^2_{\ta_3\ta_3}-
\]
\[  
  2[2\om\ta_2 + (1+3\nu) + (n - 2 M_{22})]\pa_{\ta_2} - 
\]
\begin{equation}
\label{e2.3}
 (6\om\ta_3-\frac{4}{3} M_{12} \ta_2)\pa_{\ta_3}
 +\frac{2}{3} M_{12} M_{12} - 4 \om n - 2 \om M_{22}\ .
\end{equation}
This is $n \times n$ matrix differential operator. It contains infinitely-many
finite-dimensional invariant subspaces which are nothing but finite-dimensional
representation spaces of the algebra $gl(3)$.
This operator remains exactly-solvable with the {\it same} spectra
as the scalar Calogero operator.
% but with different degeneracy (?).

Probably, this operator remains completely integrable. A non-trivial integral
is the differential operator of the sixth order $(\om \neq 0)$ or of the third order $(\om = 0)$
takes the algebraic form after the gauging away the ground state function in $\ta$ coordinates.
It can be rewritten in terms of the $gl(3)$-algebra generators of the representation $[k,0]$ which then can be replaced by the generators of the representation $[k,n]$. Under such a replacement the spectra of the integral remains unchanged and algebraic.

\section{Extension of the 3-body Sutherland Model}

The first {\it algebraic} form for the 3-body Sutherland Hamiltonian \cite{Sutherland:1971}
appears after gauge rotation with the ground state function, separation of the center-of-mass
and change the variables to elementary symmetric polynomials of the exponentials of translationally-symmetric coordinates \cite{RT:1995},
\[
        h_{\rm Suth} =
        -(2\eta_2+ \frac{\al^2}{2}\eta_2^2 -\frac{\al^4}{24}\eta_3^2) \pa_{\eta_2\eta_2}^2
\]
\[         
        -(6+\frac{4\al^2}{3}\eta_2) \eta_3\pa_{\eta_2\eta_3}^2
        +(\frac{2}{3}\eta_2^2-\frac{\al^2}{2}\eta_3^2)
                \pa_{\eta_3\eta_3}^2\ -      
\]        
\begin{equation}
\label{e2.4}        
    \bigl[2(1+3\nu)+2(\nu+{1\over 3})\al^2\eta_2\bigr]\pa_{\eta_2}
        -2(\nu+\frac{1}{3})\al^2\eta_3\pa_{\eta_3} \ ,
\end{equation}
where $\al$ is the inverse radius of the circle on where the bodies are situated.
These new coordinates are fundamental trigonometric invariants of the $A_2$ Weyl group.

As shown in \cite{RT:1995}, the operator (\ref{e2.4}) can be rewritten in a
{\it Lie-algebraic} form in terms of the $gl(3)$-algebra generators of the representation
$[k,0]$,
\[
  h_{\rm Suth} \ =\  -2E_{11} T_1^- -6 E_{22} T_1^-
\]
\[  
+\frac{2}{3} E_{12} E_{12} -2(1+3\nu)T_1^- +\frac{\al^4}{24}
E_{21} E_{21}
\]
\begin{equation}
\label{e2.5}
 -\frac{\al^2}{6}\biggl[3 E_{11} E_{11} + 8 E_{11}
E_{22} +
 3 E_{22} E_{22} + (1+12\nu)(E_{11} + E_{22})\biggr]
  \ .
\end{equation}
Now we can substitute the generators of the representation $[k,n]$
in the form (\ref{gl3})
\[
  {\tilde h}_{\rm Suth}= -(2\eta_2+{\al^2\over 2}\eta_2^2
        -\frac{\al^4}{24}\eta_3^2)\pa_{\eta_2\eta_2}^2
        -(6+\frac{4\al^2}{3}\eta_2)
        \eta_3\pa_{\eta_2\eta_3}^2
\]
\[
     +(\frac{2}{3}\eta_2^2-\frac{\al^2}{2}\eta_3^2)\pa_{\eta_3\eta_3}^2   
        -2\bigl[(1+3\nu)+(\nu+\frac{1}{3})\al^2\eta_2+(n -2 M_{22}) \bigr]\pa_{\eta_2}
\]
\begin{equation}
\label{e2.6}
   +\frac{\al^4}{24} M_{21}\eta_3\pa_{\eta_2} +
 [2(\nu+\frac{1}{3})\al^2\eta_3-\frac{4}{3} M_{12}\eta_2]\pa_{\eta_3} 
\end{equation}
\[
 -\frac{\al^2}{3}\biggl[3 n (\eta_2\pa_{\eta_2}+\eta_3\pa_{\eta_3}) +
 M_{11}\eta_3\pa_{\eta_3} + M_{22}\eta_2\pa_{\eta_2}  \biggr]
\]
\[ 
  +
 \frac{2}{3} M_{12} M_{12} +\frac{\al^4}{24} M_{21} M_{21}
\]
\[
-\frac{\al^2}{6}\biggl[ 2 M_{11} M_{22} + (1+12\nu+3n)n\biggr]\ .
\]
This is $n \times n$ matrix differential operator. It contains infinitely-many
finite-dimensional invariant subspaces which are nothing but finite-dimensional
representation spaces of the algebra $gl(3)$.
This operator remains exactly-solvable with the {\it same} spectra
as the scalar Sutherland operator.

Probably, the operator (\ref{e2.6}) remains completely integrable. A non-trivial
integral is the differential operator of the third order takes the algebraic form after
the gauging away the ground state function in $\eta$ coordinates.
It can be rewritten in terms of the $gl(3)$-algebra generators of the representation
$[k,0]$ which then can be replaced by the generators of the representation $[k,n]$.
Under such a replacement the spectra of the integral remains unchanged and algebraic.

\section*{\large Conclusions}

The algebra $gl_n$ of differential operators plays a role of the hidden algebra for {\it all}
$A_n, B_n, C_n, D_n, BC_n$ Calogero-Moser Hamiltonians, both rational and trigonometric, with the Weyl symmetry of classical root spaces (see \cite{Turbiner:2013} and references therein). We described a procedure, which to our opinion should carry the name of the {\it Havlicek procedure}, to construct the algebra $gl_n$ of the matrix differential operators. The procedure is based on a mixed, matrix-differential operators realization of the Gauss decomposition diagram.

As for the Hamiltonian reduction models with the exceptional Weyl symmetry group $G_2, F_4, E_{6,7,8}$, both rational and trigonometric, there exist hidden algebras of differential operators (see \cite{Turbiner:2013} and references therein).
All these algebras are infinite-dimensional but finitely-generated. For generating elements
of those algebras an analogue of the Weyl-Cartan decomposition exists but the Gauss decomposition
diagram: a commutator of the lowering and raising generators is a polynomial of the higher-than-one order in the Cartan generators. Surely, matrix realizations of these algebras exist. The first attempt to construct such a realization was made in Section 4.
Thus, the above mentioned procedure of building the mixed representations can be realized. It may lead to a new class of matrix exactly-solvable models with exceptional Weyl symmetry.

{\bf Acknowledgements (AVT)}

The present text was planned long ago to be dedicated to Miloslav Havlicek
who always caused for both authors a deep respect as the scientist and
the citizen.

The text is based mainly on the notes jointly prepared by the first
and the second author. It does not include the results of the authors
obtained separately (except for Section 4) and which the authors had no chance to discuss. 
Thus, it looks incomplete.
When the first author (YuFS) passed away it took years for the second author (AVT)
to return to the subject due to a sad memory. Even now, almost a decade after the death of Yura Smirnov, the preparation of this text was quite difficult for AVT.

AVT thanks CRM, Montreal for their kind hospitality extended to him where a part of this 
work was done during his numerous visits. AVT is grateful to J C Lopez Vieyra for the 
interest to the work and technical assistance.
This work was supported in part by the University Program FENOMEC, by the PAPIIT
grant {\bf IN109512} and CONACyT grant {\bf 166189}~(Mexico).

\begingroup\raggedright
\endgroup

\end{document}